\documentclass[twocolumn,showpacs,preprintnumbers,amsmath,amssymb]{revtex4}
\renewcommand\ln{\log}
\newcommand\ba{\begin{eqnarray}}
\newcommand\ea{\end{eqnarray}}

\usepackage{graphicx}
\usepackage{dcolumn}
\usepackage{bm}

\begin{document}


\title{
Lowest order QED radiative corrections \\
to longitudinally polarized
M{\o}ller scattering}

\author{A. Ilyichev}
 \email{ily@hep.by}
\affiliation{%
National Scientific 
and Educational Centre of Particle and High Energy Physics of the Belarusian State University, 
220040  Minsk,  Belarus
}%

\author{V. Zykunov}
\email{zykunov@gstu.gomel.by}
\affiliation{
Gomel State Technical University, 246746 Gomel,  Belarus
}%


\begin{abstract}
The total lowest-order electromagnetic radiative corrections to the
observables in M{\o}ller scattering of longitudinally polarized electrons
have been calculated. 
The final expressions obtained by the covariant method for the 
 infrared divergency cancellation are free from
any unphysical cut-off parameters. 
Since the calculation is carried out within the
ultrarelativistic approximation our result has a compact form
that is convenient for computing. Basing on these expressions the FORTRAN code
MERA has been developed. Using this code the detailed numerical analysis
performed under SLAC (E-158) and JLab kinematic conditions has shown
that the radiative corrections are significant and
rather sensitive to the value 
of the missing mass (inelasticity) cuts.
\end{abstract}

\maketitle

\section{Introduction}
The present intense interest of physicists in polarized M{\o}ller scattering
is stimulated by several reasons.
Today the measurement of the parity-violating asymmetry $A_{PV}$
in the recent experiment E158 \cite{E1581,E1582} at SLAC gives the
$\sin{\theta_W}$ with the best precision at low enegries.
Further, experimentally M{\o}ller scattering is actively used
in polarimetry to measure the polarization of the electron
beams \cite{polarimetry} as well as monitoring of luminosity
(for example, at DESY \cite{DESY}).
Yet another reason stimulating an interest in
M{\o }ller scattering consists in possibility to test the
Standard Model and to reveal traces of new physics.
In the intensively
discussed projects of the ILC, $e^-e^-$
and $\mu^-\mu^-$ colliders \cite{heush}, high hopes for the discovery of
Higgs bosons, manifestations of contact interactions,
the compositeness of the electron, new gauge bosons,
etc., are pinned on the scattering of identical polarized
fermions ($e,\mu$).

A precise comparison of the experimental results with the
theoretical predictions requires to take into account 
the radiative effects correctly on both QED and electroweak
levels.

Generally within the polarimetry measurements by the M{\o}ller scattering, 
the value of the transferred momentum is rather low, and therefore
electroweak effects can usually be neglected. 
But otherwise in the
projects of ILC where the energies are characterized by the TeV region,
the value of the weak and electromagnetic effects will have the same order. 
Similarly, due to specific character of
observables (singly polarized parity-violating asymmetry), 
such experiments as E-158 is sensitive to both
electromagnetic and electroweak radiative corrections (and new physics
phenomena at the TeV scales). 

The exact calculation of the lowest-order electromagnetic radiative 
corrections to polarized M{\o}ller scattering was performed by Shumeiko and Suarez
\cite{suarez}. 
The electroweak radiative corrections to polarized M{\o}ller
scattering at high energies were computed in \cite{denn} (without
hard bremsstrahlung contribution), and at low energies corresponding
to conditions of E-158 were computed in papers \cite{cz-marc}
(without hard bremsstrahlung) and \cite{petr,zyk} (including hard
bremsstrahlung).
The detailed calculation  presented there demonstrates the significant value 
of the radiative effects that have to be explicitly included both in QED and
the Electroweak theory predictions.

In the given paper we considered the lowest order electromagnetic radiative 
corrections
both to the longitudinally polarized cross sections and 
the doubly-polarized parity conserving asymmetry  
\begin{equation}
A_{LR} =
 \frac{\sigma_{LR}-\sigma_{LL}}
      {\sigma_{LR}+\sigma_{LL}},
\label{A}
\end{equation}
where the first (second) cross section subscripts $L$ and $R$
correspond left and right degree of beam (target) polarization respectivelly.
Similarly to \cite{suarez}, we perform our calculations
within the covariant Bardin--Shumeiko approach \cite{covar,shum1}, that allows to
cancel out the infrared divergences in such a way that the final result does
not depend on any unphysical parameters (such as a frame-dependent cutoff
$\Delta E$ that separates the soft photon contribution region from the hard
one). Using the ultrarelativistic approximation allows us to obtain the
compact form for radiative correction expression
that is convenient (and sometimes necessary) for fast and more
precise computer treatment. Moreover during the numerical estimations it was found that
the numerical result strictly depends on missing mass cuts. 
At the same time it should be stress that 
the first correct application of the some kinematical cuts
within the covariant Bardin--Shumeiko approach was presented in \cite{diff,ex}.         
Unfortunately Ref. \cite{suarez}
did not investigate the effects of experimentally-motivated
kinematical cuts.

The paper is organized as follows. In the Section II the kinematics of
M{\o}ller scattering as well as the cross section and asymmetry
at the lowest order are introduced.
In the Section III the  structure of the
lowest order radiative corrections  (virtual and real photon
contributions) is explained. The Section IV presents numerical
results applied to the kinematics of E-158 (SLAC) and JLab experiments.
The Section V contains some conclusions. The explicit expressions
for finite part of the real photon emission could be find in Appendix.

\section{The lowest order contribution}
\begin{figure}
\vspace*{-10mm}
\hspace*{-12mm}
\scalebox{0.27}{\includegraphics{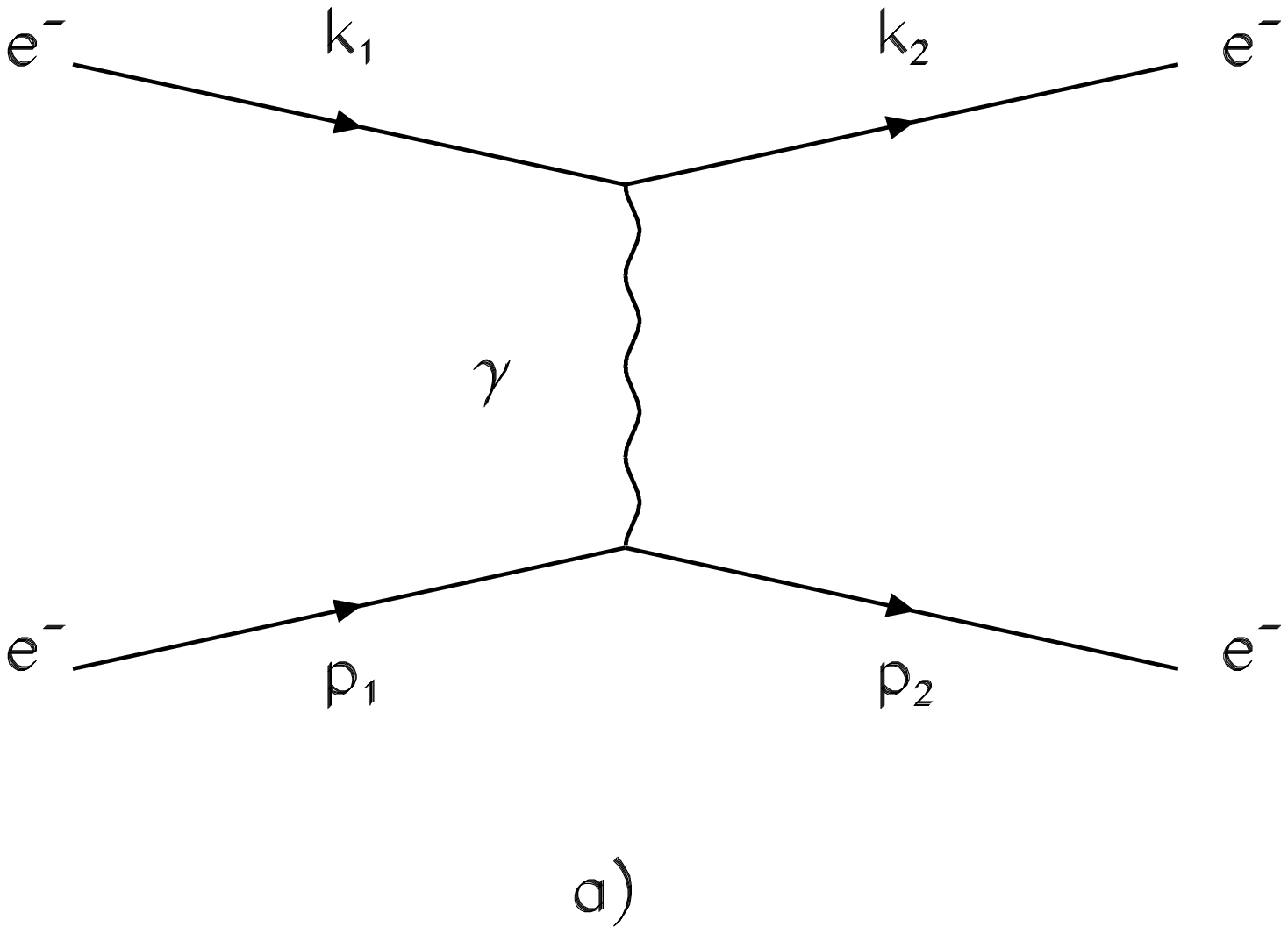}}
\hspace*{-13mm}
\scalebox{0.27}{\includegraphics{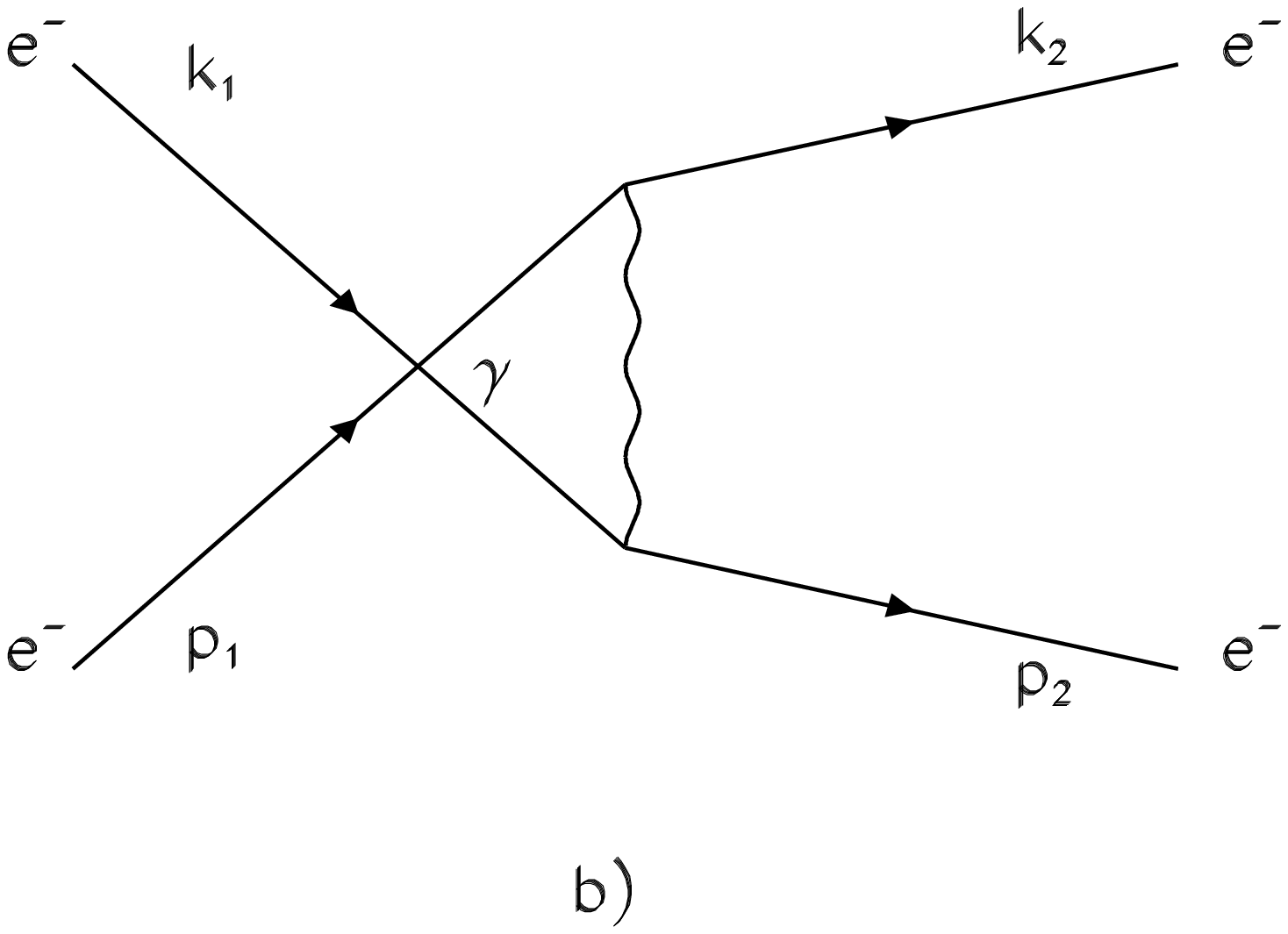}}
\vspace*{-1cm}
\caption{\label{fig:born} 
The lowest order graphs giving contribution
to the M{\o }ller scattering:
a) $t$-channel; b) $u$-channel.
}
\end{figure}

The lowest order Feynman graphs giving the contribution to M{\o}ller scattering
\ba
e(k_1,\xi_L)+e(p_1,\eta_L)\to 
e(k_2)+e(p_2) 
\ea
 are presented in Fig.~\ref{fig:born}. Here 
$k_1$, $p_1$ ($k_2$, $p_2$) are
the 4-momenta of the incoming (outgoing) electrons ($k_1^2=k_2^2=p_1^2=p_2^2=m$)
while the beam ($\xi_L$) and target  ($\eta_L$) 
polarization vectors read:
\ba
\xi_L=
\frac 1{\sqrt{s(s-4 m^2)}}
\left (\frac{s-2 m^2}mk_1-2 mp_1\right ),
\nonumber \\
\eta_L=
\frac 1{\sqrt{s(s-4 m^2)}}
\left (2 mk_1-\frac{s-2 m^2}mp_1\right ).
\ea  
Then $s$ and other Mandelstam variables can be introduce in the 
standard way:
\ba
&&s=(k_1+p_1)^2,\ t=(k_1-k_2)^2,\ u=(k_2-p_1)^2,
\nonumber \\ 
&&s+t+u=4 m^2.
\label{m1}
\ea
Notice that for the Born kinematics (strictly speaking for
the non radiative process)
\begin{equation}
u = u_0 \equiv 4m^2-s-t.
\end{equation}

Neglecting the electron mass,
the Born cross section for the M{\o}ller scattering
of longitudinal electrons
can be written as follows
\begin{equation}
\sigma^0=\frac{2\pi \alpha^2}{t^2}
\left[  (1+P)\frac{u^2}{s} - (1-P)\frac{s^2}{u}\right  ]
 + (t \leftrightarrow u).
\label{cs0}
\end{equation}
Here and later each $\sigma $ denotes the differential cross 
section over the kinematic variable $y$ 
($\sigma \equiv d\sigma /dy $) that is defined as
\begin{equation}
y=-\frac{t}{s},
\end{equation}
$P=P_BP_T$, where
$P_B, P_T$ are the polarizations of the beam and target electrons.

The form of the Born cross section (\ref{cs0}) with factorized combinations
$ 1 \pm P_BP_T $ is very convenient for construction of
the polarization asymmetry (\ref{A}) that
does not depend on any energies: 
\begin{equation}
A_{LR}^0 =
 \frac{y(1-y)(y^2-y+2)}
      {(1+y(y-1))^2}=
  \frac{\sin^2 \theta ( 7+\cos^2 \theta )}{ (3 +\cos^2 \theta )^2},
\label{A0}
\end{equation}
where $\theta$ is a scattering angle of the detected electron with
4-momentum $k_2$ in the center mass system of the initial particles 
$\vec{k}_1+\vec{p}_1=0$. The cosine of this angle can be express
via invariants in the standard way: 
\ba
\cos \theta^0=1+2t/s=1-2y,
\label{ctborn}
\ea
while the energy of the scattering lepton in Lab. system reads:
\ba
E_{k_{2}}^0=\frac{s+t-2m^2}{2m}.
\label{en0}
\ea

\section{Electromagnetic radiative corrections} 
The lowest order radiative corrections to M{\o}ller scattering
 appears from the graphs with the additional virtual particle
(V-contribution, see Fig.~\ref{fig:V} for the $t$-channel) and from the
real photon bremsstrahlung (R-contribution, see Fig.~\ref{fig:R} for the $t$-channel).
It should be noted that both these parts include the infrared divergency but their
sum must be infrared free. In this section the explicit expression 
for V- and R- contributions as well as their
infrared free sum are presented.

\subsection{Virtual contribution}

For the calculation of one-loop electromagnetic radiative corrections
we apply the on-shell renormalization scheme of electroweak standard model.
The building blocks needed for explicit calculations according this
scheme have been worked out in paper of
B\"ohm et al. \cite{BSH86}, where we take the results for gauge boson self-energies
and vertex functions.

\begin{figure}
\vspace*{-10mm}
\hspace*{-12mm}
\scalebox{0.27}{\includegraphics{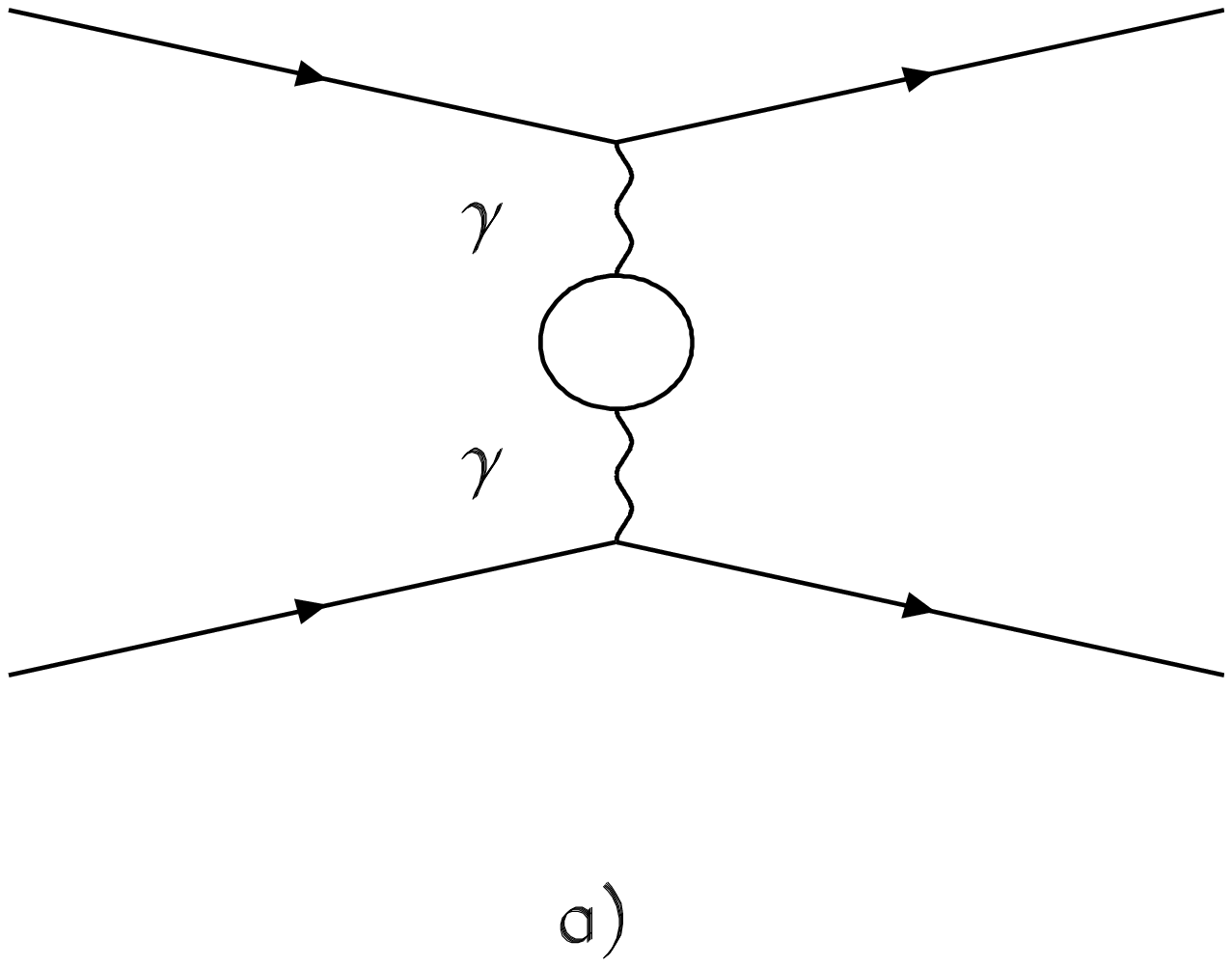}}
\vspace*{-20mm}
\hspace*{-13mm}
\scalebox{0.27}{\includegraphics{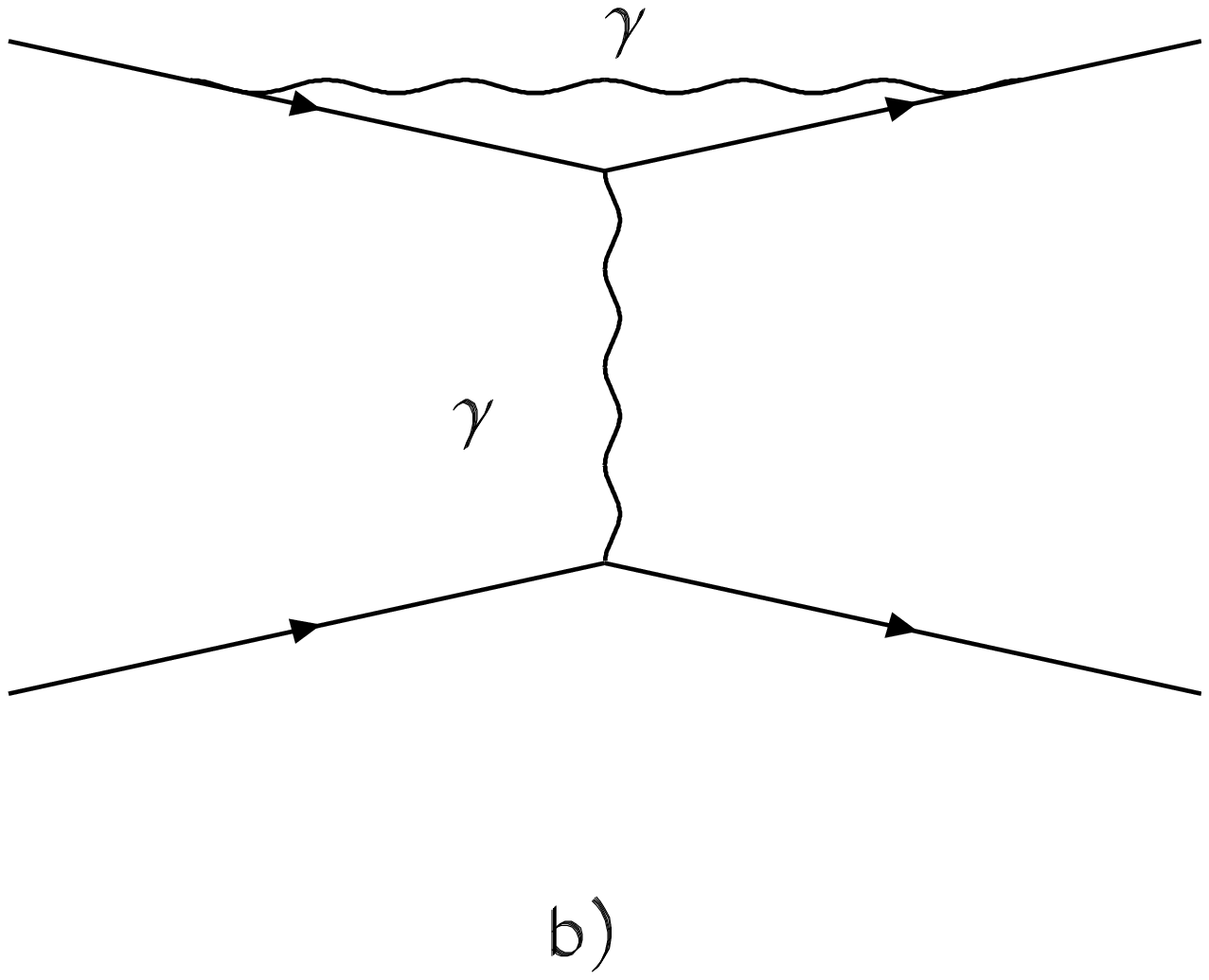}}
\hspace*{-12mm}
\vspace*{-20mm}
\scalebox{0.27}{\includegraphics{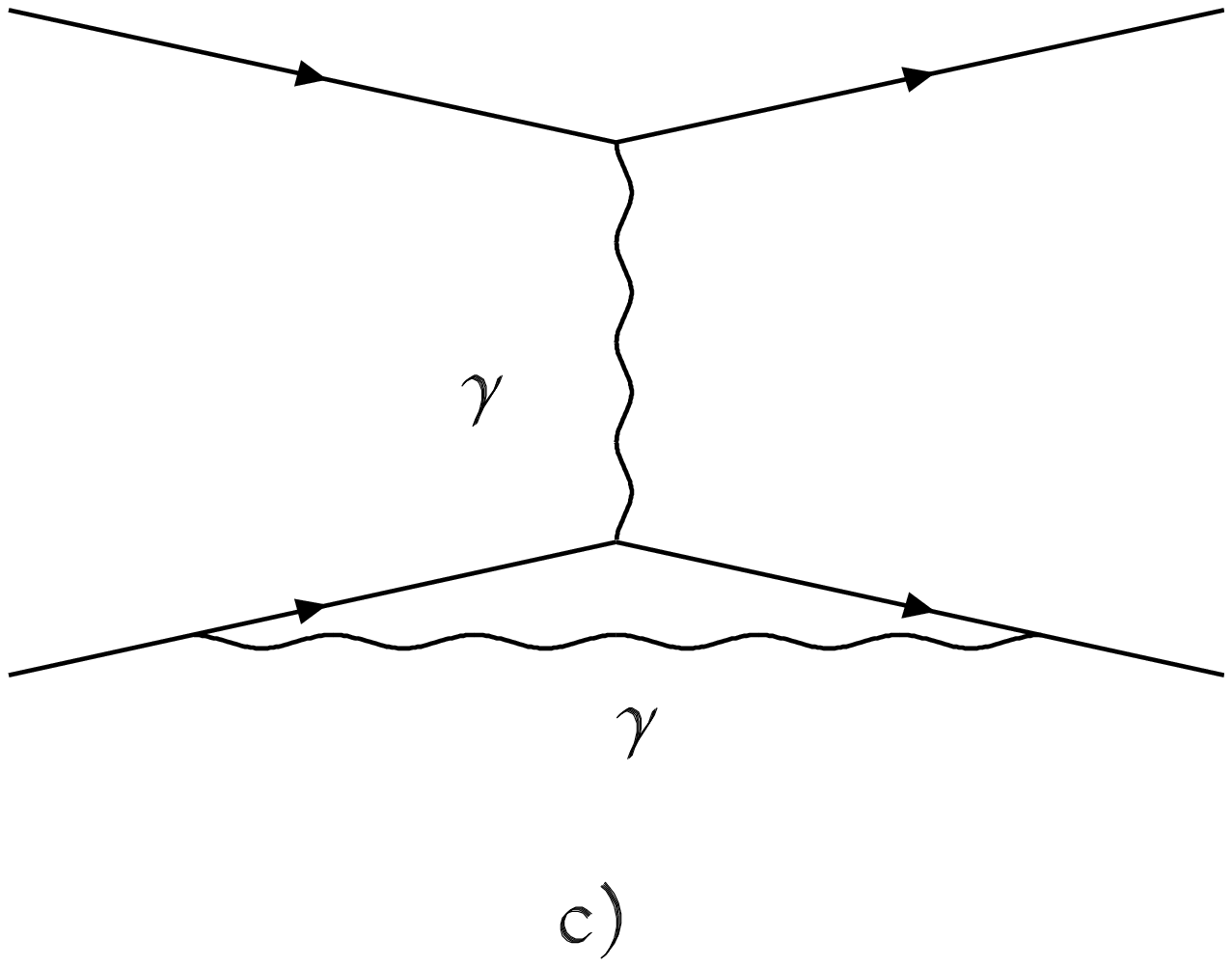}}
\hspace*{-13mm}
\scalebox{0.27}{\includegraphics{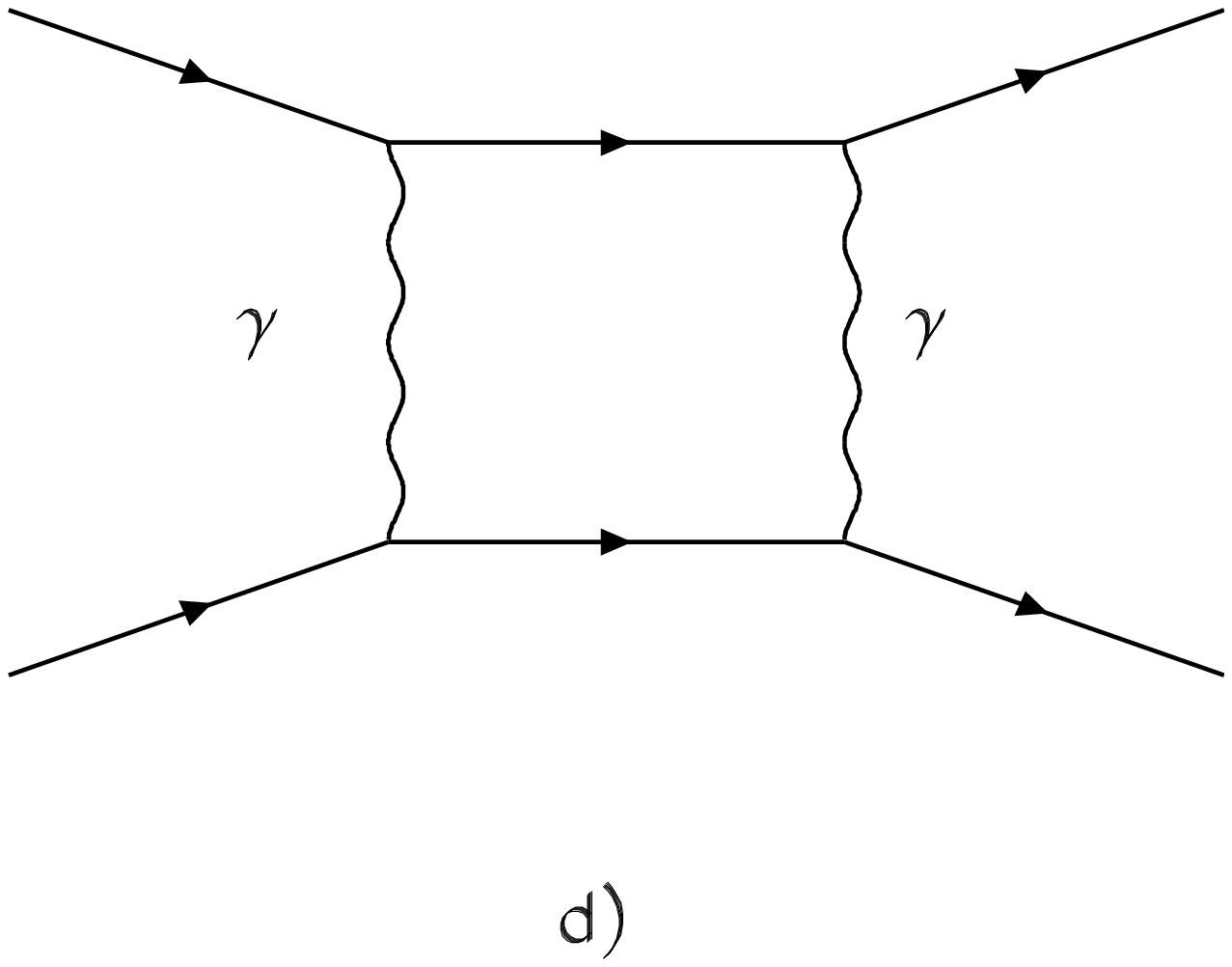}}
\hspace*{-12mm}
\scalebox{0.27}{\includegraphics{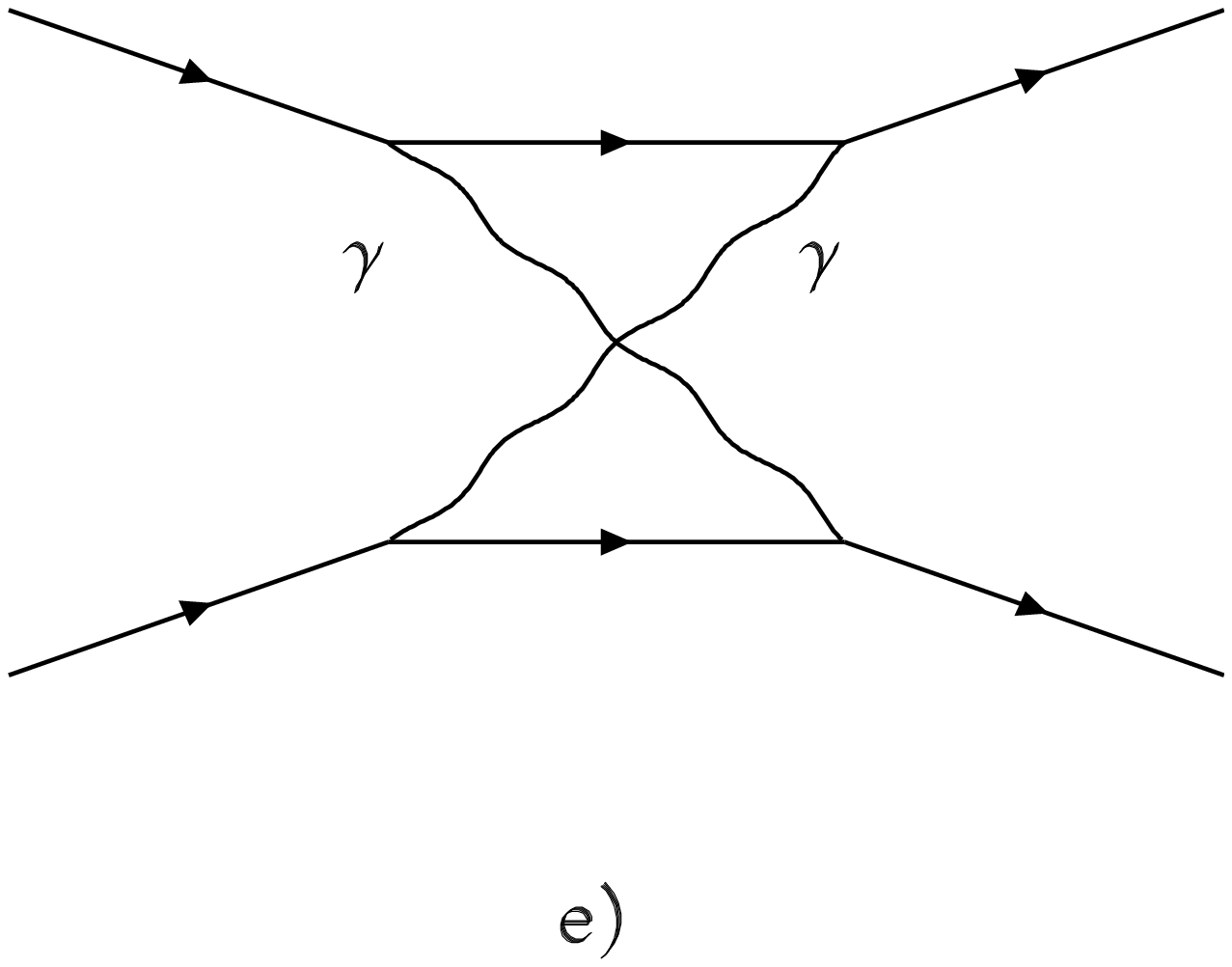}}
\vspace*{-1cm}
\caption{\label{fig:V} 
The  virtual one-loop graphs giving contribution
to the corrected M{\o }ller scattering
within  $t$-channel.
}
\end{figure}

The virtual contributions to M{\o}ller scattering
can be separated into
three parts:
\begin{equation}
\sigma^{V}=
\sigma^S+\sigma^{Ver}+\sigma^{Box},
\label{G}
\end{equation}
where
\begin{enumerate}
\item
$\sigma^S$ is a virtual photon self-energy contribution [Fig.~\ref{fig:V} (a)];
\item
$\sigma^{Ver}$ is a vertex function  contribution [Fig.~\ref{fig:V} (b,c)];
\item
$\sigma^{Box}$ is a box  contribution [Fig.~\ref{fig:V} (d,e)].
\end{enumerate}

Now we consider each of them.
\begin{enumerate}
\item
The contribution of the virtual photon self energies 
(including the photon vacuum polarization by hadrons)
to the cross section looks like
\begin{eqnarray}
\sigma^S&=& \frac{4\pi \alpha^2}{t^2}
\mbox{Re} \Bigl( -\frac{1}{t} \hat{\Sigma}^{\gamma}_T(t)
 + \Pi_h(-t) \Bigr )
\times
\nonumber \\[0.3cm] \displaystyle
&&
\times
\Bigl[ 
 (1+P)\frac{u^2}{s}
   -(1-P)\frac{s^2}{u}  \Bigr ]
 + (t \leftrightarrow u).
\label{self-en}
\end{eqnarray}
Here $\hat{\Sigma}^{\gamma}_T(-t)$ is the renormalized transverse
part of the $\gamma$--self-energy \cite{BSH86}
(this part includes
vacuum polarization by  $e$, $\mu$ and $\tau$ charged leptons:
in corresponding formula of \cite{BSH86} we should take a
summing index $f=e,\mu,\tau$).
Hadronic part of the photonic vacuum polarization associated with
light quarks can be directly obtained from the data on
process $e^+e^- \rightarrow \mbox{hadrons}$ via dispersion relations.
Here we use parameterization of \cite{Bur-Piet}
\begin{equation}
\mbox{Re} \Pi_h(-t) \cong A+B \ln (1+C|t|),
\end{equation}
with updated parameters A,B,C in different energy regions.

\item
For the contribution of  electron vertices we used the
results of the paper \cite{BSH86} (see also references therein).
We can obtain the vertex part as
\begin{eqnarray}
\sigma^{Ver}&=&\frac{2 \alpha^3}{t^2}
\Bigl[  (1+P)\frac{u^2}{s}
\nonumber \\   &&
   -(1-P)\frac{s^2}{u}  \Bigr] \Lambda_1(t,m^2)
 + (t \leftrightarrow u),
\label{csv}
\end{eqnarray}
where
\begin{eqnarray}
\Lambda_1(t,m^2) &=& -2\ln\frac{|t|}{\lambda^2} 
\Bigl(\ln\frac{|t|}{m^2}-1\Bigr)+
\nonumber \\[0.3cm] \displaystyle
&&+ \ln\frac{|t|}{m^2} + \ln^2\frac{|t|}{m^2} + 4(\frac{\pi^2}{12}-1).
\end{eqnarray}
\item
Our calculation for the box cross section gives compact formula:
\begin{eqnarray}
\sigma^{Box}&=&
\frac{2 \alpha^3}{t}
\Bigl[   \frac{1+P}{s}
\Bigl( \frac{2u^2}{t}\ln\frac{s}{|u|}\ln\frac{|su|}{\lambda^2m^2}
\nonumber \\[0.3cm] \displaystyle
&& -\delta^1_{(\gamma\gamma)} \Bigr)-  \frac{1-P}{u}
       \Bigl( \frac{2s^2}{t}\ln\frac{s}{|u|}\ln\frac{|su|}{\lambda^2m^2}
\nonumber \\[0.3cm] && 
         -\delta^2_{(\gamma\gamma)} \Bigr ) \Bigr ]
 + (t \leftrightarrow u),
\end{eqnarray}
The expressions $\delta_{(\gamma\gamma)}^{1,2}$ have the form:
\begin{eqnarray}
\delta^1_{(\gamma \gamma)} & = &
l_s^2\frac{  s^2+u^2}{2t} - l_su -(l_x^2+\pi^2)\frac{u^2}{t},
\nonumber \\[0.3cm] \displaystyle
\delta^2_{(\gamma \gamma)} &=&
  l_s^2\frac{s^2}{t} + l_x s - (l_x^2 + \pi^2 )\frac{s^2+u^2}{2t},
\end{eqnarray}
and logarithms look like
\begin{equation}
l_s=\ln\frac{s}{|t|},\ l_x=\ln\frac{u}{t}.
\end{equation}
\end{enumerate}
It should be noted that vertex and box parts contain the infrared divergence
through the appearance of the fictitious photon mass $\lambda$.
The infrared part from virtual cross section
can be extracted in a simple way:
\begin{eqnarray}
\sigma^{V}_{IR} &&=
\sigma^{V} - \sigma^{V}(\lambda^2 \rightarrow s)
\nonumber \\[0.3cm] \displaystyle
&&= -\frac{2\alpha}{\pi} \ln\frac{s}{\lambda^2}
  \left( \ln\frac{tu}{m^2s}-1 \right) \sigma^{0}.
\end{eqnarray}

\subsection{Real bremsstrahlung contribution}

\begin{figure}
\vspace*{-10mm}
\hspace*{-12mm}
\scalebox{0.27}{\includegraphics{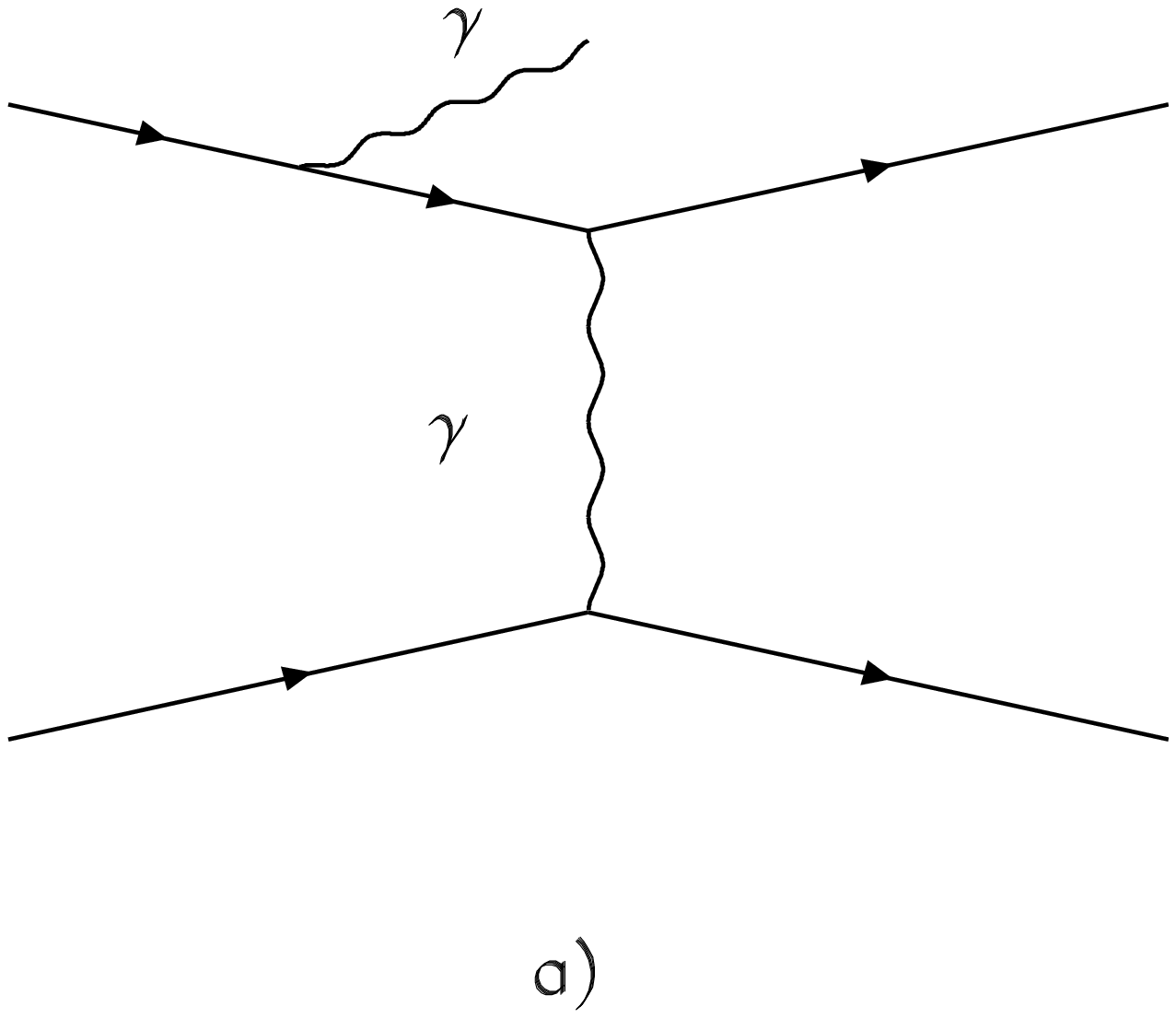}}
\vspace*{-20mm}
\hspace*{-13mm}
\scalebox{0.27}{\includegraphics{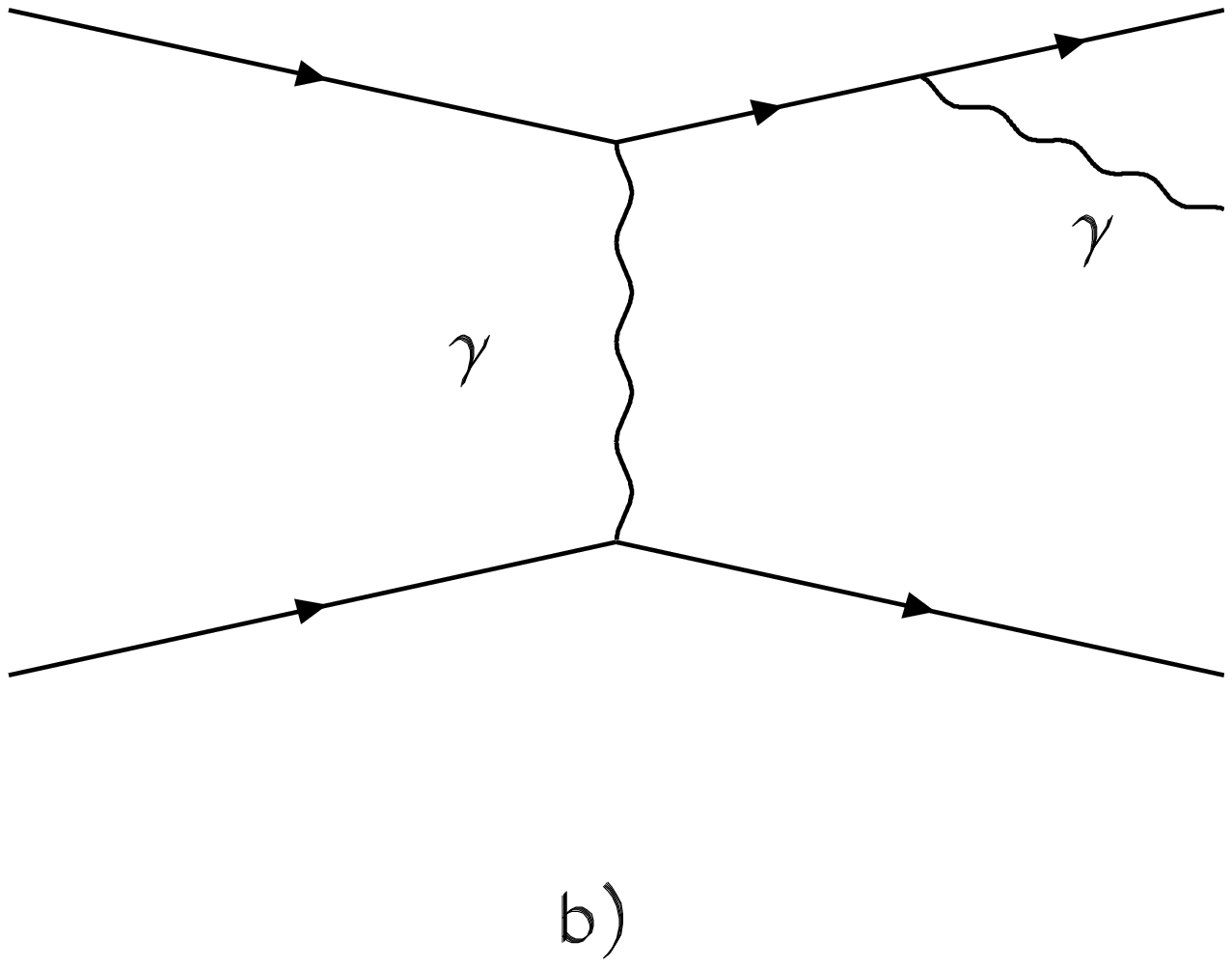}}
\hspace*{-12mm}
\vspace*{-1.5cm}
\scalebox{0.27}{\includegraphics{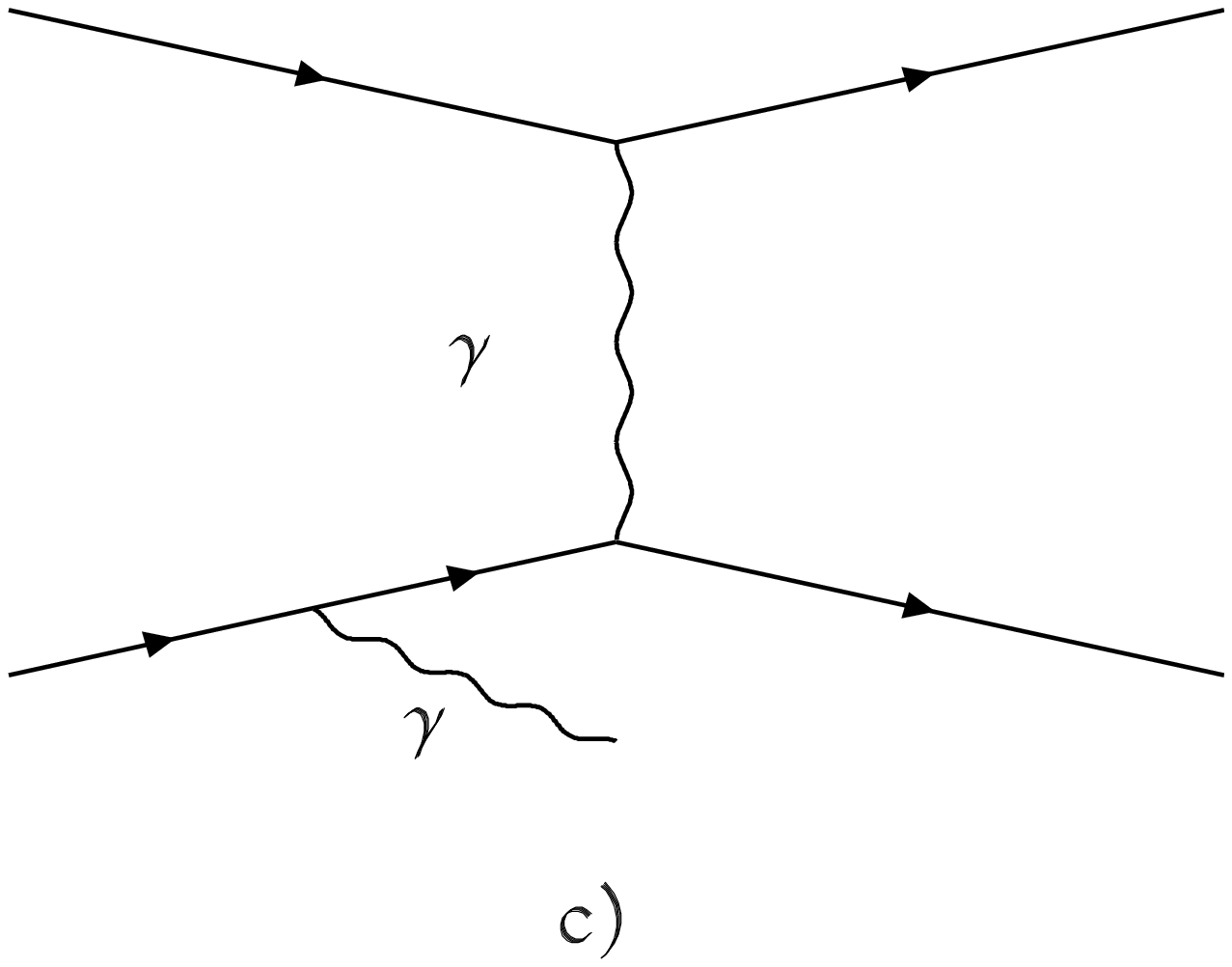}}
\hspace*{-13mm}
\scalebox{0.27}{\includegraphics{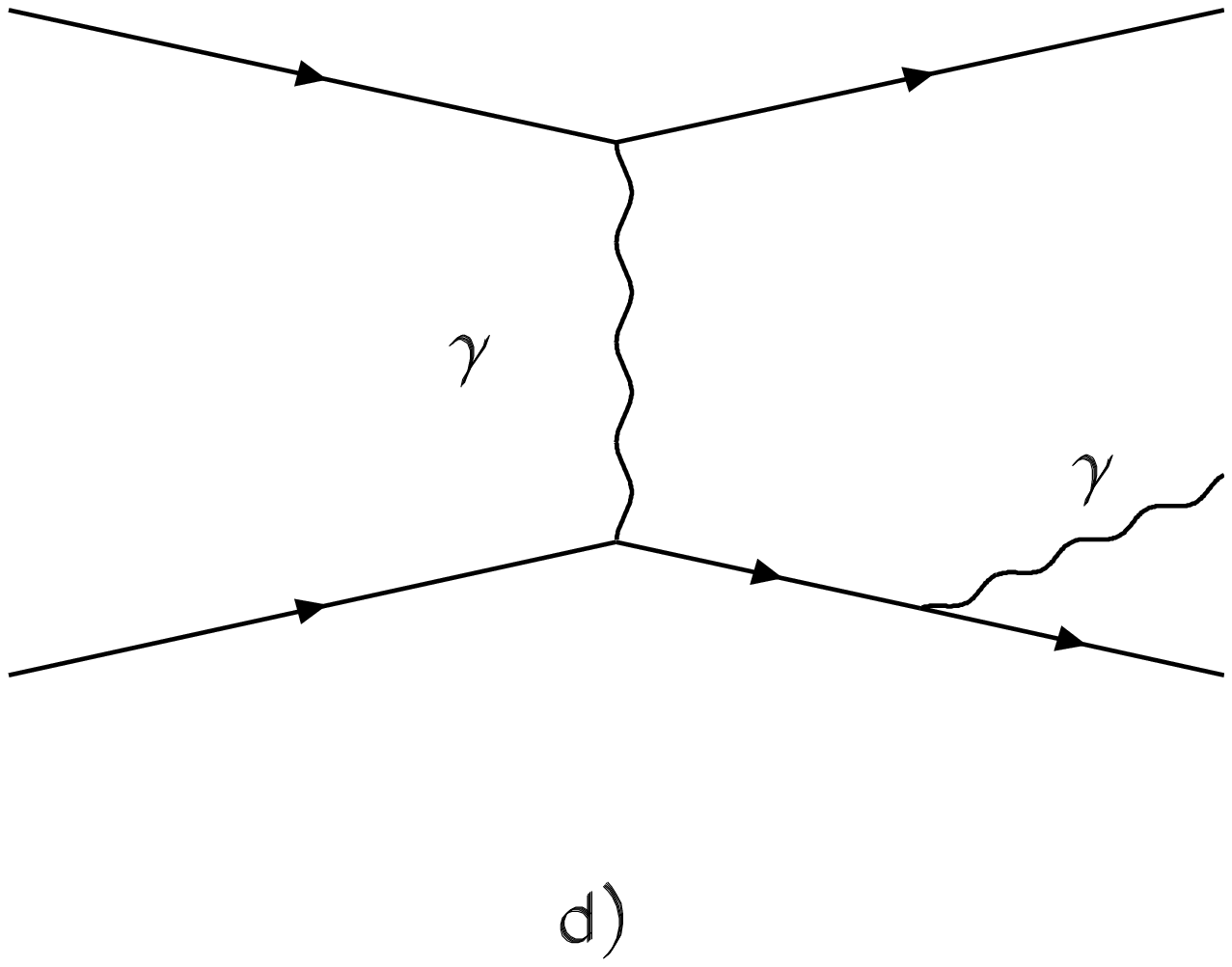}}
\vspace*{5mm}
\caption{\label{fig:R}
The real one photon emission graphs giving contribution
to the corrected M{\o }ller scattering
within $t$-channel. 
}
\end{figure}

The full set of Feynman graphs contributed to 
the real photon bremsstrahlung are presented in Fig.~\ref{fig:R}.
For extraction of the infrared divergency 
we use the prescription of Bardin and Shumeiko \cite{covar}:
\ba
\sigma^R\equiv
\sigma^R-\sigma^R_{IR}+\sigma^R_{IR}=
\sigma^R_{F}+\sigma^R_{IR},
\ea
where the infrared free part can be presented in the following way
\begin{equation}
\sigma^R_F
= - \frac{\alpha^3}{\pi s}
   \int\limits_0^{v_{max}}dv
\sum_{i=1}^{10} S_i.
\label{srf} 
\end{equation}
The explicit expressions for $S_i$ are presented in the Appendix.
The integration in (\ref{srf}) is performed over variable
$v$ that is a so-called inelasticity.
The reason of this term can be explain by the fact 
that for the radiative process 
the last relation in (\ref{m1}) transforms into
\ba
&&s+t+u=v+4 m^2.
\label{m2}
\ea
The explicit expression for $v$ can be defined as
$v=\Lambda^2-m^2$, where $\Lambda=k_1-k_2+p_1$ and $\Lambda^2$ is
a so-called missing mass squared.

It should be noted that due to kinematical restrictions 
the upper limit of the integration in (\ref{srf}) is defined as 
\ba
v_{max}=\frac{st+\sqrt{s(s-4m^2)t(t-4m^2)}}{2m^2}\sim s+t. 
\ea
On the other hand, the energy of the scattering lepton in Lab. system
(\ref{en0}) for the radiative process transforms into 
\ba
E_{k_{2}}^R=\frac{s+t-v-2m^2}{2m},
\label{enr}
\ea
\\and reaches its minimum value for $v=v_{max}$
\ba
E_{k_{2}}^R|_{v=v_{max}}=\frac{s+t-v_{max}-2m^2}{2m}\sim -m\frac{s^2+t^2}{2st}.
\label{enrm}
\ea
Obviously the electron with the energy (\ref{enrm}) cannot be detected. 
Moreover, as it was point out first in \cite{diff}, the variable $v$ can be 
directly reconstructed from the measured momenta. However
not all events with non-zero $v<v_{cut}$ can be rejected 
from the experimental data due to finite resolution of the experimental equipment.  
Therefore during the radiative corrections calculation for the given 
experimental setup it is necessary to take into account this fact. 

Notice that for the radiative events the cosine of the scattering angle 
in the centre mass system of the initial particles also depends 
on integration variable $v$:
\ba
\cos \theta^R=1+2t/(s-v).
\label{ctrad}
\ea

The infrared-divergent part of brems\-strah\-lung cross section
integrated over the real photon phase space is given in terms of a finite
(and infinitesimal) photon mass $\lambda$ in
\begin{equation}
\sigma_{IR}^R=
\frac{\alpha}{\pi}
\left[ 4 \ln \frac{v_{{max}}}{m\lambda}
  \left (\ln \frac{tu}{m^2s}-1\right) + \delta_1^S + \delta_1^H
\right ] \sigma^0,
\label{ir}
\end{equation}
where  (see \cite{covar} for details) 
\ba
\delta_1^S &=& -\frac 1 2l_m^2+(3-2 l_r)l_m
-(l_m-1)\ln \frac{s(s+t)}{t^2}-\frac 12l_r
\nonumber \\[0.3cm]
&&-\frac{\pi ^2}{3}+1
\ea
and 
\begin{figure*}
\vspace*{-10mm}
\hspace*{-5mm}
\scalebox{0.45}{\includegraphics{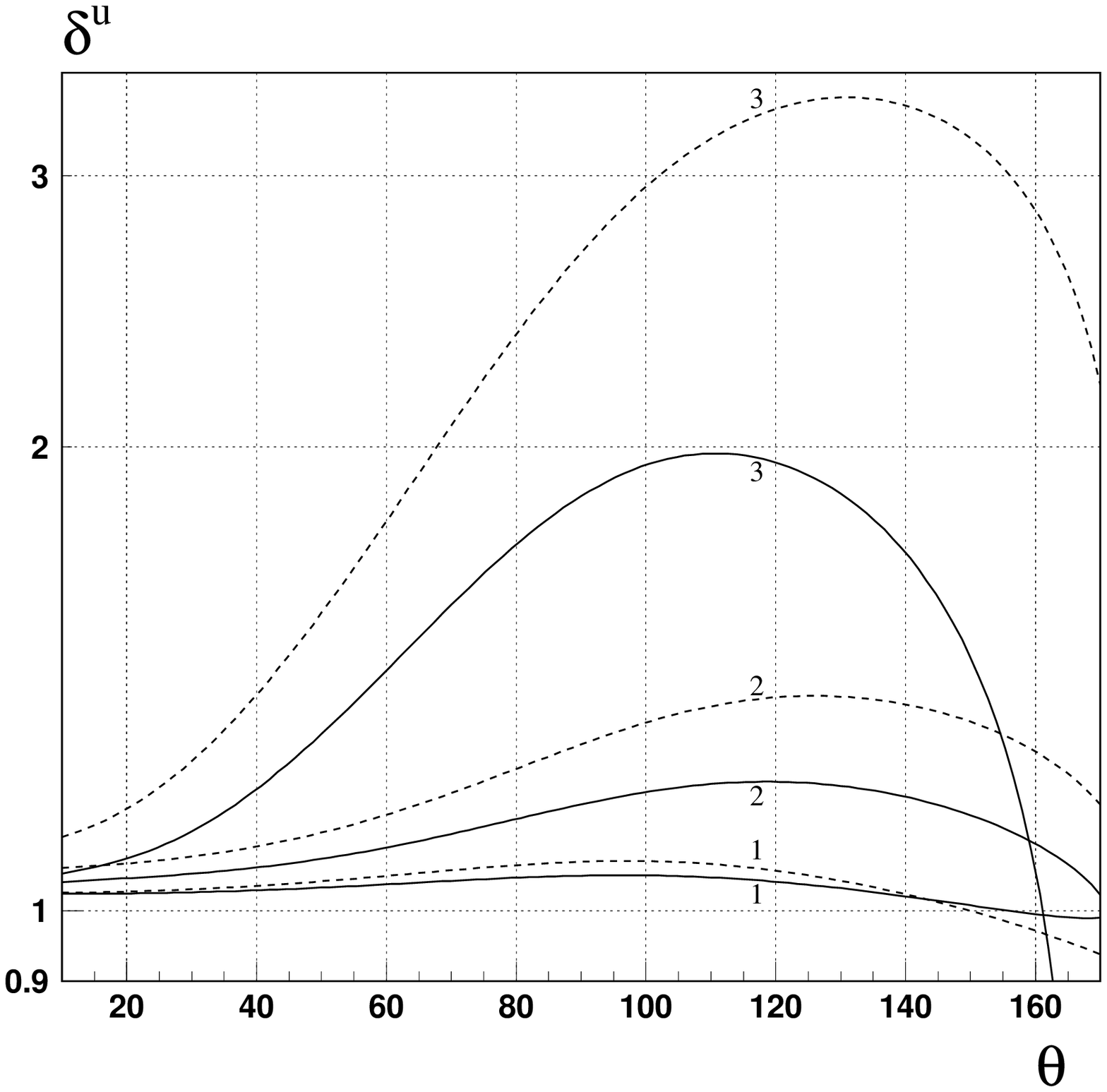}}
\scalebox{0.45}{\includegraphics{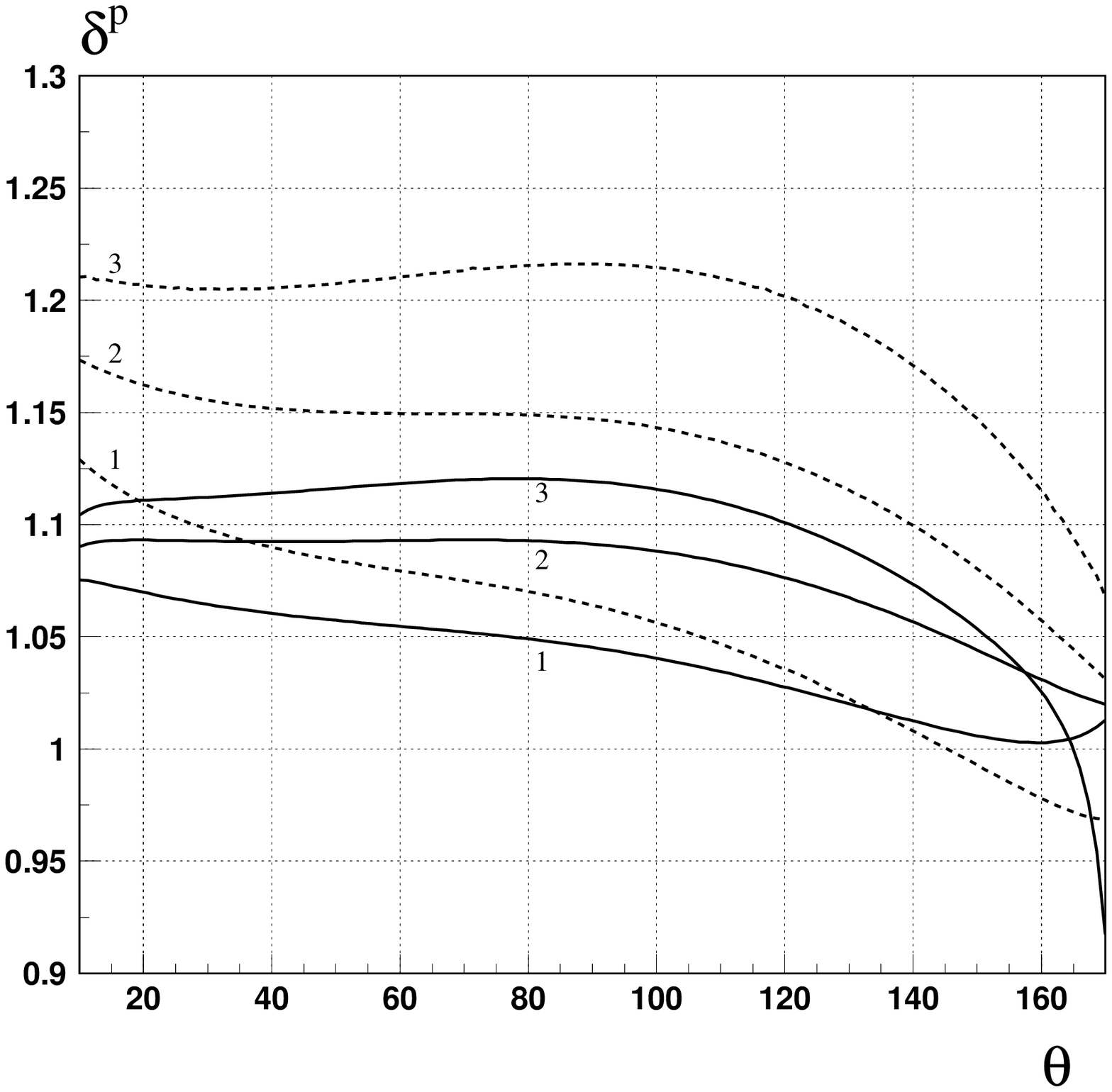}}
\vspace*{-5mm}
\caption{\label{fig:1} 
The relative corrections to the unpolarized ($\delta ^u$) and polarized 
($\delta ^p$) parts of the cross section as a functions of the scattering
angle (defined according to eq. (\ref{ctapp})) for JLab 
 ($E_{beam}=1$ GeV, solid lines) and SLAC 
($E_{beam}=45$ GeV, dashed lines) kinematic conditions 
with different inelasticity cuts: 
1) $v_{cut}=0.5v_{max}$; 2) $0.9v_{max}$; 3) $0.99v_{max}$. 
}
\end{figure*}
\vspace*{-15mm}
\begin{widetext}
\ba
\delta_1^H  & = &-\frac 5 2l_m^2+
\Bigl (\ln \frac{t^2(s+t)^2(s-v_{max})}{s(s+t-v_{max})^2v_{max}(v_{max}-t)}
+1\Bigr )l_m-\frac 1 2 \ln ^2\frac{v_{max}}{|t|}
-\ln ^2\bigl(1-\frac{v_{max}}{t}\bigr)
\nonumber \\[0.2cm]&&
+\ln \frac {s+t}{s+t-v_{max}}
\ln \frac {(s+t)(s+t-v_{max})}{t^2}
+\ln \frac {s-v_{max}}{|t|}
\ln \frac {s-v_{max}}{s}
+\ln \frac {v_{max}}{|t|}
\nonumber 
\ea
\ba
\nonumber\\[-0.6cm]
\;\;\;\;\;\;
\;\;\;\;\;\;
\;\;\;\;\;\;
\;\;
+2\Bigl[
{\rm Li}_2\left(\frac{v_{max}}s\right)
-{\rm Li}_2\left(\frac{v_{max}}t\right)
-{\rm Li}_2\left(\frac{v_{max}}{s+t}\right)\Bigl]
+{\rm Li}_2\left(\frac{s-v_{max}}s\right)
-{\rm Li}_2\left(\frac{t-v_{max}}t\right)
-\frac{\pi^2}6.
\label{dh1}
\ea
\end{widetext}
Here ${\rm Li}_2(x)$ is the Spence function and  
\ba
&& l_m=\ln\frac{-t}{m^2},\ l_r=\ln\frac{s+t}{s}.
\ea

Summing up (\ref{G}) and (\ref{ir})
\begin{eqnarray}
\sigma^{RV} &=&
\sigma^{R}_{IR}+
\sigma^{V}=
\frac{\alpha}{\pi}
( 4 \ln \frac{v_{{max}}}{m\sqrt{s}}
  (\ln \frac{tu}{m^2s}-1) + \delta_1^S 
\nonumber \\[0.3cm] \displaystyle
&&
+ \delta_1^H) 
\sigma^0+
\sigma^{V}(\lambda^2 \rightarrow s),
\label{can}
\end{eqnarray}
 we obtain a cancellation of infrared divergencies 
 from R- and V- contribution.

Finally, the total infrared free radiative corrected cross
section reads:
\ba
\sigma^{obs}=
\sigma^{0}+
\sigma^{RV}+
\sigma^{R}_F.
\label{obs}
\ea

\section{Numerical estimations}

Basing on the equations presented above the FORTRAN code 
MERA\footnote{FORTRAN code MERA is available from
http:/www.hep.by/RC} 
(M{\o}ller scattering: Electromagnetic RAdiative corrections) has 
been developed. 
In this section using MERA the
numerical estimation of radiative effects to the  
M{\o}ller scattering of longitudinally polarized electrons
is presented. 
 
There are two basic differences between the numerical analysis
that are performed in this and previous
\cite{suarez,zyk} papers: we  show the dependence 
of radiative corrections  on the scattering angle in the center mass 
system of the initial electrons, we investigate  
the dependence of radiative corrections on the value of the missing mass cut.

As it was mentioned above for the radiative events the cosine of 
the scattering angle has to be expressed not only via $t$ and $s$ as for 
non-radiative events (\ref{ctborn}) but and via inelasticity 
$v$ too (see (\ref{ctrad})).
Taking into account that we calculate the cross section as a function of $y$ 
or $t=-ys$ (because $s$ is fixed) and inelasticity is the inegration variable 
we has some uncertanties in the definition of 
the scattering angle for the observable cross section (\ref{obs}).
In order to escape it we use the standard non-radiative  
approximation (\ref{ctborn}), i.e. we assume that
\ba
\cos \theta^R\sim
\cos \theta^0
=1+2t/s.
\label{ctapp}
\ea
\begin{figure*}
\vspace*{-10mm}
\hspace*{-5mm}
\scalebox{0.45}{\includegraphics{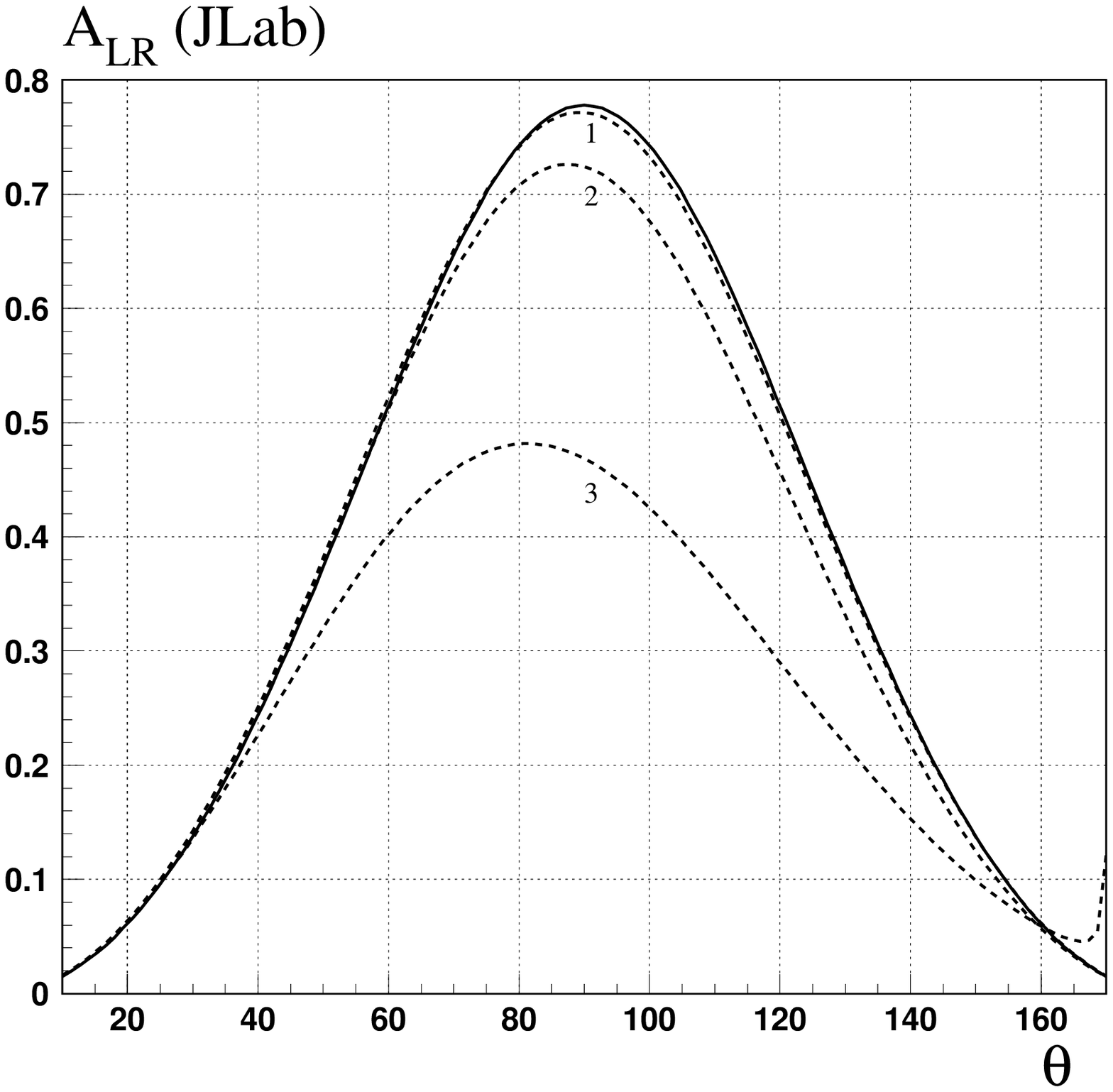}}
\scalebox{0.45}{\includegraphics{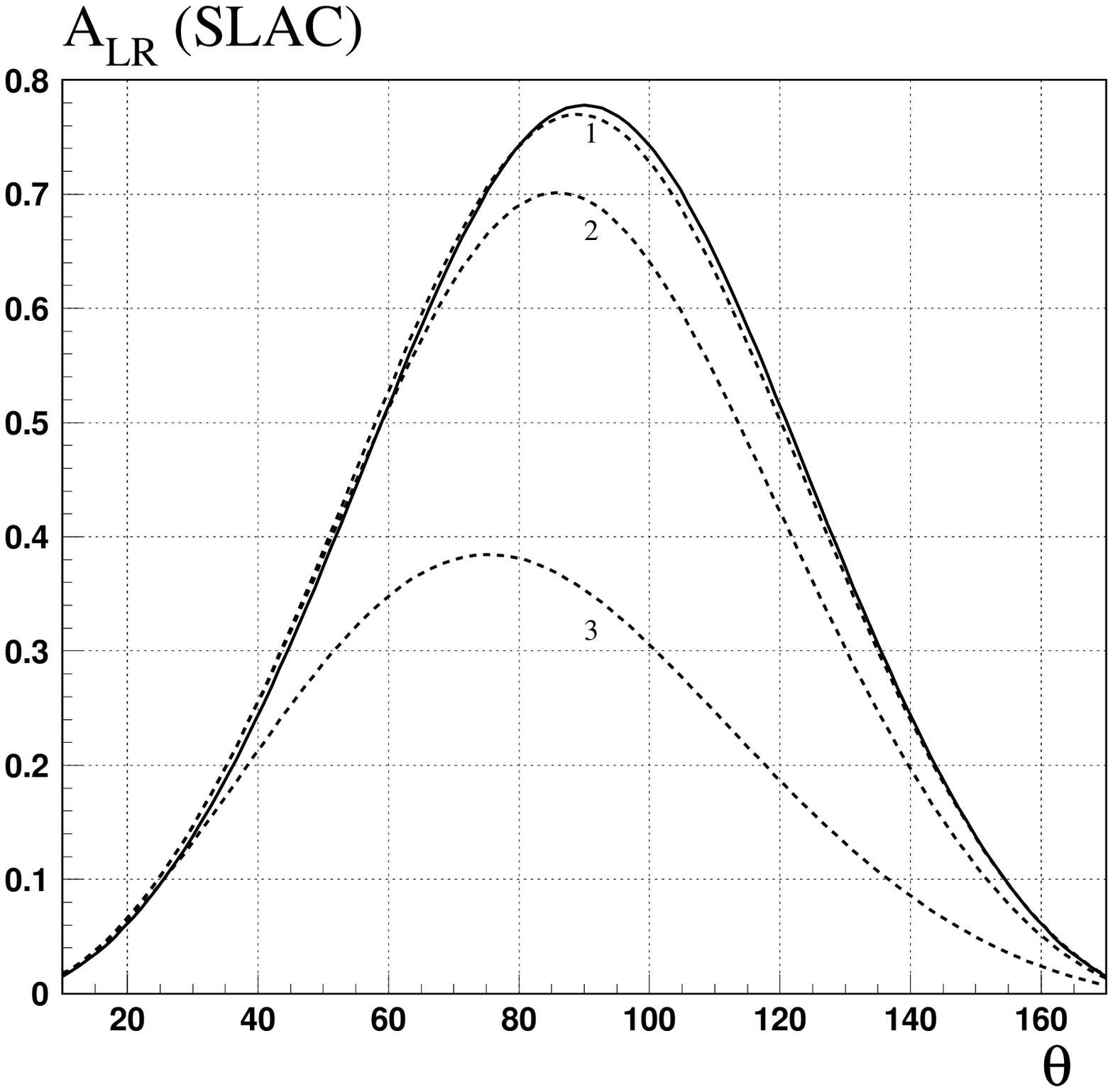}}
\vspace*{-5mm}
\caption{\label{fig:2}
$\theta$-dependence ($\theta$ is defined according to eq. (\ref{ctapp}))
of the Born (solid line)
and observable (dashed lines) asymmetries 
for JLab ($E_{beam}=1$ GeV) and SLAC 
($E_{beam}=45$ GeV) kinematic conditions 
with different inelasticity cuts: 
1) $v_{cut}=0.5v_{max}$; 2) $0.9v_{max}$; 3) $0.99v_{max}$. 
}
\end{figure*}
\begin{figure}
\vspace*{-10mm}
\hspace*{-5mm}
\scalebox{0.45}{\includegraphics{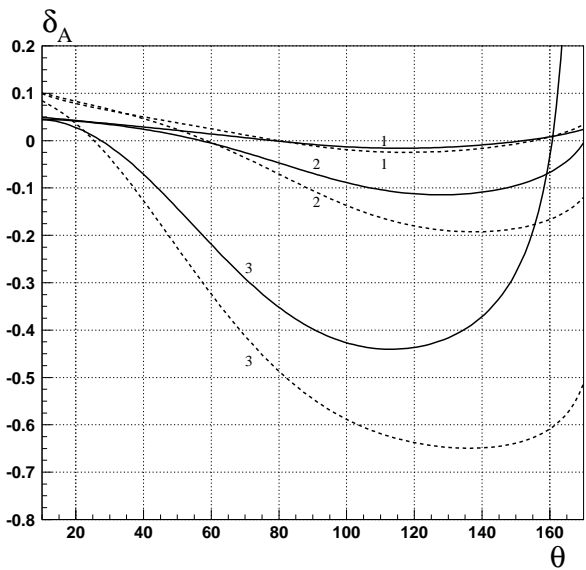}}
\vspace*{-1cm}
\caption{\label{fig:1qq}
The relative corrections to the asymmetry (\ref{da}) as a functions of the scattering
angle  (defined according to eq. (\ref{ctapp})) for JLab ($E_{beam}=1$ GeV, 
solid lines) and SLAC 
($E_{beam}=45$ GeV, dashed lines) kinematic conditions 
with different inelasticity cuts: 
1) $v_{cut}=0.5v_{max}$; 2) $0.9v_{max}$; 3) $0.99v_{max}$. 
}
\end{figure}

The cross section for polarized M{\o}ller scattering can be 
presented as a difference of the unpolarized and polarized parts
\ba
\sigma^{0,obs}=
\sigma^{0,obs}_u-P
\sigma^{0,obs}_p.
\ea
In the Fig.~\ref{fig:1} the $\theta$-dependence of the relative 
radiative correction for the unpolarized and polarized  parts
of the cross section 
\ba
\delta ^{u,p}=\sigma^{obs}_{u,p}/\sigma^{0}_{u,p}
\label{das}
\ea
for three
different inelasticity  cuts  $v_{cut}$ 
(and, therefore $\Lambda ^2$ cuts): 
$v_{cut}=0.5v_{max}$, $0.9v_{max}$ and $0.99v_{max}$ is presented.
One can see the following features of their behavior:
the presence of maximum values at $\theta \geq 90^0$; sizable 
increasing when $v_{cut}$ tending to its maximum value. For the 
 $v_{cut}=0.5v_{max}$ the corrections $\delta ^u$ ($\delta ^p$) 
 for $\theta=90^0$ equal to 1.075 (1.064) for SLAC and 1.054 (1.045) for JLab.

The $\theta $-dependence of the Born and observable asymmetries 
with the same inelasticity cuts is presented  in Fig.~\ref{fig:2}. 
In this figure it can be seen that in the most region of $\theta $
the corrected asymmetries are less then the Born ones and essentially
decrease with the increasing $v_{cut}$. 
The Fig.~\ref{fig:1qq}, where $\theta $-dependence of
the relative corrections to the asymmetries 
\ba
\delta_A=\frac{A^{obs}_{LR}-A^{0}_{LR}}{A^{0}_{LR}}
\label{da}
\ea
is presented, reflects this fact clear. Particularly, it could be seen
that for the realistic $v_{cut}=0.5v_{max}$ 
we have $\delta_A=-0.008 (-0.01)$
for JLab (SLAC) at $\theta=90^0$ while for $v_{cut}=0.99v_{max}$
the relative corrections to the asymmetries reach 
$\delta_A=-0.4 (-0.55)$ for JLab (SLAC) at the same angle.

Let us consider the situation with more realistic
small $v_{cut}$. As it could be
seen from (\ref{dh1}) and (\ref{can}) the radiative corrected cross section
(\ref{obs}) diverge when $v_{cut}$ tends to zero. Such cross section
behavior can be explain in a simple way. Naturally that
there is no any real photon emission in the limit $v_{cut}\rightarrow 0$. Therefore we need
to say about the infrared divergency that appear from V-contribution and can not be 
canceled due to any real photon emission absent.  

\begin{figure}
\vspace*{-10mm}
\hspace*{-5mm}
\scalebox{0.45}{\includegraphics{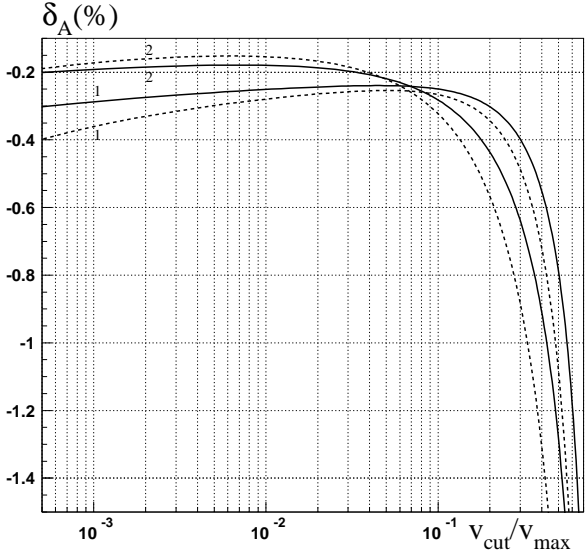}}
\vspace*{-1cm}
\caption{\label{fig:1qq2}
The relative corrections to the asymmetry (\ref{da}) as a functions 
of $v_{cut}/v_{max}$ at the scattering
angle (defined according to eq. (\ref{ctapp})) in CM system 1) $\theta =90^0$; 2) $100^0$
for JLab ($E_{beam}=1$ GeV, solid lines) and SLAC 
($E_{beam}=45$ GeV, dashed lines) kinematic conditions. 
}
\end{figure}

The other very interesting feature  consists in the deviation 
of the observable asymmetry from the Born one at the small $v_{cut}$
where asymmetry reach its maximum value. Due to rather small
effects once again in Fig.~\ref{fig:1qq2} the quantity (\ref{da}) as a function of the ratio 
$v_{cut}/v_{max}$ is presented for JLab and SLAC kinematic conditions. From this picture it 
can be seen that for $0.001v_{max}<v_{cut}<0.1v_{max}$ 
the relative correction to the asymmetry is flat 
and consists some dozen of percent while starting with $v_{cut}>0.1v_{max}$ it
rapidly falls. 


\section{Conclusions}

The explicit expressions for the lowest order electromagnetic
radiative corrections to M{\o}ller scattering of 
longitudinally polarized
electrons in ultrarelativistic approximation have been obtained.
Basing on these expressions the FORTRAN code MERA has been developed.

The numerical analysis performed for different values of
missing mass (inelasticity) cut has shown that
the radiative corrections are strongly depended on this
parameter.
So when this cut tends to its maximum
value two tendencies should be observed:
the radiatively corrected cross sections increase while
the radiatively corrected asymmetries decrease. At the same
time $\theta $-dependence has shown some common features:
the relative corrections to the cross sections (asymmetries)
has a maximum (minimum) at $\theta \geq 90^0$.
For more realistic small cuts the relative corrections to the asymmetry are
rather flat and amounts some dozen percent.

Taking into consideration a large scale of obtained 
radiative effects we proof the necessity of radiative
correction procedure for JLab and SLAC experiments. 
Particularly to perform  data processing correctly
it is necessary to construct Monte Carlo generator
for simulation of radiative events within M{\o }ller
scattering.

\begin{acknowledgments}
The authors would like to thank Andrei Afanasev, 
Igor Akushevich, Eugene Chudakov,
Yury Kolomensky, Nikolai Shumeiko and Juan Suarez  
for stimulating discussions. We also thank
Nikolai Shumeiko and Juan Suarez for agreement to compare 
the numerical results from FORTRAN codes M{\O}LLERAD and MERA.
V.Z. (A.I.) would like to thank SLAC (JLab) staff for their generous hospitality
during their visits.
\end{acknowledgments}

\appendix*

\section{}
$S_1$ is the $t$-channel contribution of the emission from
the upper electron leg:
\widetext
\begin{eqnarray}
S_1 &=& S_1^u+PS_1^p + S_1^{a},\
\nonumber \\[0.3cm] \displaystyle
S_1^u &=& - L_A - \hat L_A
+ L_s(2s^2t^{-2} + 2t^{-1}u + 2t^{-2}u^2 + 1)
+ L_x( - 2s^2t^{-2} - 2st^{-1} - 2t^{-2}u^2 - 1)
- 2L_t + 4L_m
\nonumber \\[0.3cm] &&
- 2t^{-2}( v(s^2+u^2)s^{-1}u^{-1}-2t ),
\nonumber \\[0.3cm] \displaystyle
S_1^p &=&
- L_A + \hat L_A
+ L_s( - 2t^{-1}u - 1)
+ L_x( - 2st^{-1} - 1)
+ 4L_mt^{-1}(s - u) + 4t^{-2}(s + t - u)
\nonumber \\[0.3cm] \displaystyle
&&
-2(4s^2+2st-t^2)s^{-1}t^{-1}(s+t)^{-1}
-2(3s+2t)t^{-1}u^{-1}+2us^{-1}t^{-1},
\nonumber \\[0.3cm] \displaystyle
S_1^{a} &=& 4(1+P)t^{-2}(u+u_0)(1+tL_m);
\end{eqnarray}
%
%
%
$S_2$ is the $t$-channel contribution of the interference from
the upper and lower electron legs:
\begin{eqnarray}
S_2 &=& (S_2^u+PS_2^p)/t + S_2^{a},\
\nonumber
\end{eqnarray}
\begin{eqnarray}
S_2^u &=&
L_s (8 s^3 t^{-1} - 8 s^2 t^{-1} v + 8 s^2 + 4 s t + 4 s t^{-1} v^2
- 6 s v
- t^2 + 2 t v - v^2)/(t - v)
+ L_x ( - 8 s^3 t^{-1} + 16 s^2 t^{-1} v
\nonumber \\[0.3cm] \displaystyle
&&
- 16 s^2 - 12 s t
- 12 s t^{-1} v^2 + 22 s v - 5 t^2 + 12 t v
+ 4 t^{-1} v^3 - 11 v^2)/(t - v)
+ 2 L_u (2 s + t - v)
\nonumber \\[0.3cm] \displaystyle
&&
+ L_1 (4 s^3 t^{-1} - 4 s^2 t^{-1} v + 2 s^2 + s t + 2 s t^{-1}
 v^2 - s v
+ t v - v^2)
+ L_2 ( - 4 s^3 t^{-1} + 8 s^2 t^{-1} v - 10 s^2 - 9 s t
\nonumber \\[0.3cm] \displaystyle
&&
- 6 s t^{-1} v^2 + 13 s v - 3 t^2 + 7 t v
+ 2 t^{-1} v^3 - 6 v^2)
+ 2 L_3 t ( - 2 s - t + v)
+ 2 t ((v-s)^{-1}-(s+t)^{-1}
\nonumber \\[0.3cm] \displaystyle
&&
+ 2 (2 s+t-v)/(t-v)^2),
\nonumber 
\end{eqnarray}
\begin{eqnarray}
S_2^p &=&
L_s (12 s^2 t + 8 s^2 t^{-1} v^2 - 20 s^2 v + 4 s t^2 - 12 s t v
- 4 s t^{-1} v^3 + 12 s v^2 - t^3 + 3 t^2 v - 3 t v^2 + v^3)
/(t-v)^2
\nonumber \\[0.3cm] \displaystyle
&&
+ L_x ( - 12 s^2 t - 8 s^2 t^{-1} v^2 + 20 s^2 v - 20 s t^2+ 52 s t v
+ 12 s t^{-1} v^3
- 44 s v^2 - 7 t^3 + 25 t^2 v - 33 t v^2
- 4 t^{-1} v^4
\nonumber \\[0.3cm] \displaystyle
&&
+ 19 v^3)
/(t-v)^2
+ 2 L_u t
+ L_1 (2 s^2 t^2 - 8 s^2 t v
- 4 s^2 t^{-1} v^3 +10 s^2 v^2 + s t^3
- 3 s t^2 v + 5 s t v^2 + 2 s t^{-1} v^4
\nonumber \\[0.3cm] \displaystyle
&&
- 5 s v^3
+ t^3 v - 3 t^2 v^2
 + 3 t v^3 - v^4)
/(t-v)^2
+ L_2 ( - 2 s^2 t^2 + 8 s^2 t v + 4 s^2 t^{-1} v^3
- 10 s^2 v^2
- 3 s t^3
\nonumber \\[0.3cm] \displaystyle
&&
+ 17 s t^2 v
- 31 s t v^2 - 6 s t^{-1} v^4 + 23 s v^3 -
 t^4 + 7 t^3 v
- 17 t^2 v^2
+ 19 t v^3
+ 2 t^{-1} v^5 - 10 v^4)
/(t-v)^2
- 2 L_3 t v
\nonumber \\[0.3cm] \displaystyle
&&
-2 t (t+v)(s+t)^{-1}(s-v)^{-1},
\nonumber 
\end{eqnarray}
\begin{eqnarray}
S_2^{a} &=& 2t^{-2}\{ (L_x-\hat L_A)(s^2(1-P)+(u^2+uu_0+u_0^2)(1+P))
+ s(L_A+L_s)(1+P)(u+u_0) \};
\end{eqnarray}
%
%
%
$S_3$ is the $t$-channel contribution of the emission from the lower electron leg:
\begin{eqnarray}
S_3 &=& (S_3^u+PS_3^p)/t^2 + S_3^{a},\
\nonumber \\[0.3cm] \displaystyle
S_3^u &=& -L_u t(s^2+u^2)(v-2t)(t-v)^{-2}
+ (3 s^2 t^3 {\tau}^{-1}
+ 12 s^2 t^2 - 13 s^2 t v + 4 s^2 v^2
+ 3 s t^4 {\tau}^{-1}
+ 9 s t^3
\nonumber \\[0.3cm] \displaystyle&&
- 25 s t^2 v + 17 s t v^2
- 4 s v^3+ \frac{3}{2} t^5 {\tau}^{-1} + t^4 - 13 t^3 v
+ 19 t^2 v^2
- \frac{21}{2} t v^3
 + 2 v^4)/(t-v)^3,
\nonumber \\[0.3cm] \displaystyle
S_3^p &=& L_u t (s-u)(2t-v)(t-v)^{-1}
 + \frac{(s-u)}{2\tau}(st^3+8t^2v
-7tv^2+4v^3)(t-v)^{-2},
\nonumber \\[0.3cm] \displaystyle
S_3^a &=& 2(1+P)t^{-2}(u+u_0)(1+tL_u);
\end{eqnarray}
%
%
%
$S_{4,5,6,7}$ are the contributions of the interference between the $t$- and $u$-channel
graphs:
\begin{eqnarray}
S_4 &=& ((1-P)S_4^c+PS_4^p)/u + S_4^a,\
\nonumber \\[0.3cm] \displaystyle
S_4^c &=&
 - 2 L_1 s^3 t^{-1} + L_2 (2 s^3 t^{-1} + 4 s^2 + 3 s t + t^2)
+ 2 \hat L_A ( - 3 s^3 - 7 s^2 t + 2 s^2 v - 5 s t^2
+ 3 s t v - t^3
\nonumber \\[0.3cm] \displaystyle
&&
+ t^2 v)
(s + t)^{-2}
+ 2 L_m (2 s - v) + L_s s t^{-1} (2 s - v)
+ L_x ( - 4 s^2 t^{-1}
+ 3 s t^{-1} v - 6 s - 3 t - t^{-1} v^2 + 3 v)
\nonumber \\[0.3cm] \displaystyle
&&
+ L_t ( - 2 s - 3 t + 2 v)
+ L_3 t (2 s + t )
+ 2 ( - 2 s^3 t^{-1} v
+ s^3 - 2 s^2 t - 3 s t^2 + 3 s t v - s v^2
\nonumber \\[0.3cm] \displaystyle
&&
      - t^2 v + t v^2)
(s+t)^{-2}(t-v)^{-1},
\nonumber \\[0.3cm] \displaystyle
S_4^p &=&
4 s^2 t^{-1} v(s + t)^{-2},
\nonumber \\[0.3cm] \displaystyle
S_4^a &=& 2s^2(1-P)( 2-u_0\hat L_A+2tL_m+sL_A )/(tuu_0);
\end{eqnarray}
\begin{eqnarray}
S_5 &=& (1-P)S_5^c + S_5^a,\
\nonumber \\[0.3cm] \displaystyle
S_5^c &=&
L_5 v
+ 2 L_6 ( - s^3 - s^2 u + s^2 v - s v^2 - u v^2 + v^3)
(s+u)^{-1}(v-u)^{-1}t^{-1}
- \hat L_A
+ 2 L_m ( - 2 s^4 u^{-1}
\nonumber \\[0.3cm] \displaystyle
&&
+ 5 s^3 u^{-1} v - 4 s^3 - 3 s^2 u
- 6 s^2 u^{-1} v^2 + 10 s^2 v - 2 s u^2
+ 9 s u v
+ 4 s u^{-1} v^3
- 11 s v^2 - u^3 + 4 u^2 v - 6 u v^2 - u^{-1} v^4
\nonumber \\[0.3cm] \displaystyle
&&
+ 4 v^3)
(s+u)^{-1}(v-u)^{-1}t^{-1}
+ L_r ( - 2 s^3 - 2 s^2 u + s^2 v - u^2 v)
(s+u)^{-1}(v-u)^{-1}t^{-1}
\nonumber \\[0.3cm] \displaystyle
&&
+ L_s (4 s^4 u^{-1} - 4 s^3 u^{-1} v
+ 4 s^3 + 4 s^2 u
+ 3 s^2 u^{-1} v^2
- 8 s^2 v - 4 s u v
- s u^{-1} v^3 + 6 s v^2
+ 2 u v^2 - 2 v^3)
(s+u)^{-1}
\nonumber \\[0.3cm] \displaystyle
&&
(v-u)^{-1}t^{-1}
- L_t + L_u
- L_x (4 s^3 u + 2 s^3 u^{-1} v^2 - 6 s^3 v
- s^2 u^2 + 2 s^2 u v
- s^2 v^2 - 2 s u^3 + 7 s u^2 v - 8 s u v^2
+ 3 s v^3
\nonumber \\[0.3cm] \displaystyle
&&
- u^4
+ 3 u^3 v - 4 u^2 v^2 + 3 u v^3 - v^4)
(s+u)^{-1}(u-v)^{-2}t^{-1}
+ 4
(2s-v)u^{-1}t^{-1},
\nonumber \\[0.3cm] \displaystyle
S_5^a &=& 2s^2(1-P)( 2+s L_s+2tL_m+u_0L_x )
     t^{-1}u^{-1}u_0^{-1};
\end{eqnarray}
\begin{eqnarray}
S_6 &=& (1-P)S_6^c/(tu) + S_6^a,\
\nonumber \\[0.3cm] \displaystyle
S_6^c &=&
2 L_u s^2
- 3 {\tau}^{-1} s^2 t^3
(t-v)^{-3}
+ t (5 s^4 t - 2 s^4 v + 10 s^3 t^2
- 4 s^3 t v + s^2 t^3
+ 10 s^2 t^2 v
- 12 s^2 t v^2 + 4 s^2 v^3 - 4 s t^4
\nonumber \\[0.3cm] \displaystyle
&&
+ 14 s t^3 v - 18 s t^2 v^2 + 10 s t v^3 - 2 s v^4 + t^4 v
- 3 t^3 v^2
+ 3 t^2 v^3 - t v^4)
(s+t)^{-2}(t-v)^{-3}
\nonumber \\[0.3cm] \displaystyle
S_6^a &=& 2s^2(1-P)( s L_s-u_0\hat L_A+tL_u )
     t^{-1}u^{-1}u_0^{-1};
\end{eqnarray}
\begin{eqnarray}
S_7 &=& ((1-P)S_7^c+PS_7^p)/(tu) + S_7^a,\
\nonumber \\[0.3cm] \displaystyle
S_7^c &=&
- 2 L_1 s^3 + L_4 t (s^2 + t^2)
+ L_5
(s+t)(s-t-v)u
+ L_s s (2 s - v) + L_x ( - 4 s^2 - 2 s t
+ 3 s v + t v - v^2)
\nonumber \\[0.3cm] \displaystyle
&&
+ L_u ( - 3 s^2 t^2 + 4 s^2 t v - s^2 v^2 - 2 s t^3 + 6 s t^2 v
       - 6 s t v^2
+ 2 s v^3 - t^4
+ 3 t^3 v - 4 t^2 v^2 + 3 t v^3
       - v^4)
(t-v)^{-2}
\nonumber \\[0.3cm] \displaystyle
&&
+ 2 ( - 4 s^3 t + 2 s^3 v - 7 s^2 t^2 + 5 s^2 t v - 3 s t^3
+ 3 s t^2 v
- t^3 v + 2 t^2 v^2 - t v^3)
(s+t)^{-1}(t-v)^{-2},
\nonumber \\[0.3cm] \displaystyle
S_7^p &=&
4 s^{-1} v (s^3 + 2 t^2 v - 4 t v^2 + 2 v^3)(t-v)^{-2},
\nonumber \\[0.3cm] \displaystyle
S_7^a &=& 2s^2(1-P)( 2+u_0L_x+tL_u+sL_A )
     t^{-1}u^{-1}u_0^{-1};
\end{eqnarray}
$S_{8,9,10}$ are the pure u-channel contributions:
\begin{eqnarray}
S_8 = (S_3^u+PS_3^p)|_{t \leftrightarrow u}/u^2 + S_8^{a},\
S_8^a = -2(s^2+t^2-(s^2-t^2)P)((u+u_0)u^{-2}u_0^{-2}
             -\hat L_A u^{-1}u_0^{-1});
\end{eqnarray}
\begin{eqnarray}
S_9 = (S_2^u+PS_2^p)|_{t \leftrightarrow u}/u + S_9^{a},\
S_9^a = -2(s^2+t^2-(s^2-t^2)P) (2tL_m+sL_A+sL_s
+ tL_u )  (u+u_0)u^{-2}u_0^{-2};
\end{eqnarray}
\begin{eqnarray}
S_{10} = (S_1^u+PS_1^p)|_{t \leftrightarrow u} + S_{10}^{a},\
S_{10}^a = -2(s^2+t^2-(s^2-t^2)P)(2(u+u_0)u^{-2}u_0^{-2}
+ L_xu^{-1}u_0^{-1});
\end{eqnarray}

Evidently, the contributions $S_1,\ S_2,\ S_3$ 
are in agreement with the corresponding terms of calculations
\cite{old-bard-shum} (unpolarized fermion scattering)
and \cite{my1} (longitudinally polarized fermion scattering),
if we suppose that masses of the initial fermions being equal.

Here we present the logarithms (and their combinations) which
were used in hard brems\-strah\-lung calculation (notice that
all of them do not lead to infrared singularity):
\ba
&& L_m = -\frac{1}{t} \log\frac{|t|}{m^2},\
   L_A = -\frac{1}{v-s} \log\frac{(v-s)^2}{m^2\tau},
\nonumber \\[0.3cm] \displaystyle
&&
   \hat L_A = -\frac{1}{v-u} \log\frac{(v-u)^2}{m^2\tau},
   L_t = \frac{1}{v-t} \log\frac{\tau(v-t)^2}{m^2t^2},\
\nonumber \\[0.3cm] \displaystyle
&&
   L_s = \frac{1}{s} \log\frac{s^2}{m^4},\
   L_x = -\frac{1}{u} \log\frac{u^2}{m^4},\
\nonumber 
\ea
\ba
&&
   L_u = \frac{1}{v-t} \log\frac{(v-t)^2}{m^2\tau},\
   L_1 = \frac{1}{v} (L_s-L_A),\
\nonumber \\[0.3cm] \displaystyle
&&
 L_2 = \frac{1}{v} (L_x+\hat L_A),\
 L_3 = \frac{1}{v} L_t,\
\nonumber \\[0.3cm] \displaystyle
&&
   L_4 = \frac{1}{v} (L_u-2L_m),\
   L_5 = \frac{1}{v(s+t)} \log\frac{u^2m^2}{(v-u)^2\tau},\
\nonumber \\[0.3cm] \displaystyle
&&
   L_6 = \frac{1}{s} \log\frac{s\tau^2}{m^2ut},\
   L_r = \frac{1}{u} \log\frac{t^2}{u^2},\; \tau=v+m^2.
\ea
\twocolumngrid
\bibliography{ura}
\end{document}